\documentstyle[aps,epsf, psfig,preprint]{revtex}

\newcommand{\bib}{\bibitem}

\begin{document}

\draft

\title{$K_L \to \gamma \nu \bar{\nu}$ decay beyond the standard model}

\author{Ji-Hao Jiang \footnote{jjh@mail.ustc.edu.cn}, Dao-Neng Gao
\footnote{gaodn@ustc.edu.cn}, and Mu-Lin Yan
\footnote{mlyan@ustc.edu.cn}}
 \address{Center for Fundamental
Physics, and Interdisciplinary Center for Theoretical Study\\
University of Science and Technology
 of China, Hefei, Anhui 230026 China}
\maketitle

\begin{abstract}
\noindent
 The decay $K_L \to \gamma \nu \bar{\nu}$ is investigated
beyond the standard model. Interestingly, the upper limit of the
$CP$-conserving and $CP$-violating branching ratios of the decay,
induced from the possible extensions of the standard model, would be
larger than the corresponding
branching ratios given in the standard model respectively, and it is
expected that the $CP$-violating part could be enhanced.
\end{abstract}
\vfill
USTC-ICTS-02-05

Rare kaon decays play an important role in studying
flavor-changing neutral currents (FCNC) and $CP$-violating
phenomena in modern particle physics \cite{BB01}. Within the
context of the standard model, they are suggested to test the
Cabibbo-Kobayashi-Maskawa (CKM){\cite{N.Cabibbo}} paradigm.  They
also provide an ideal place to search for new physics beyond the
standard model \cite{Colangelo:1998pm,rarekaon}. Even the new
physics may appear in $B$-meson decays, it is interesting to
study the footprint it leaves in rare kaon decays and especially
the $s \rightarrow d$ weak transition.

The decays $K^+ \to \pi^+ \nu \bar{\nu}$ and $K_L \to \pi^0 \nu
\bar{\nu}$ are both governed by the same $s \to d \nu \bar{\nu}$
transition. They are dominated from the short-distance loop
contributions containing virtual heavy quarks. It is believed that
the long-distance contributions in these decays are much smaller
than the short-distance ones, and could be negligible
\cite{{D. Rein},{M. Lu},{C.Q. Geng}}. Due to the absence of
possible large theoretical uncertainty from the long-distance,
these two modes are very useful both to test the standard
model and to explore the new physics. Experimentally, the decay
$K^+ \to \pi^+ \nu \bar{\nu}$ has been measured by E787 group at
BNL \cite{Adler:2001xv}
\begin{equation}
B(K^+ \to \pi^+ \nu \bar{\nu}) = (1.57_{-0.82}^{+1.75}) \times
10^{-10},
\end{equation}
which is consistent with the prediction of the standard model
\cite{Buras:2001af}
\begin{equation}
B(K^+ \to \pi^+ \nu \bar{\nu})_{SM}=(0.72\pm0.21) \times 10^{-10}.
\end{equation}

The amplitude of $K_L \rightarrow \pi^0 \nu \bar{\nu}$ in the
standard model is proportional to the CKM factor
$Im(V^*_{td}V_{ts})$ and the decay branching ratio is found to be
at the level of $10^{-12}$ \cite{Belanger:1990ur}. Whereas the
current experiment limit is less than $5.9\times
10^{-7}$\cite{Alavi-Harati:1999hd}, and this work is ongoing. But
from the experimental point of view, it is difficult to find the
event, because all the final-state particles are neutral, only the
$2\gamma$'s from $\pi^0$ can be detectable.

An alternative way is to search the decay $K \to \pi\pi \nu
\bar{\nu}$ \cite{Geng:1996kd,LV96,Chiang:2000bg}. However, the
decay branching ratio is small and the background for $ \pi\pi$ is
large . So another good choice is to study the decay of $K_L \to
\gamma \nu \bar{\nu}$\cite{MRM,Geng:2000ny}, where there is only
one photon at the final state.

The total branching ratio of the decay $K_L\to\gamma \nu\bar{\nu}$
can be divided into CP-conserving part and CP-violating part
\cite{Geng:2000ny}. The contribution within the standard model to
this decay has been investigated by some authors, and it is
believed that the mode is short-distance dominated
\cite{MRM,Geng:2000ny}. The purpose of this paper is to examine
the possible contribution to this decay from the new physics
scenarios beyond the standard model. It will be shown below that
there exists the significant model-independent limit, from which
the branching ratio of $CP$-conserving part could be up to the
same order of magnitude as the one predicted within the standard model
while the $CP$-violating part could be of more than one order of
magnitude lager than the one given in the standard model. As
an interesting example, we also consider the possible dominant
supersymmetric contributions to this decay mode.

The effective Hamiltonian for $s \rightarrow d \nu \bar{\nu}$
transition generally takes the form

\begin{eqnarray}
H_{eff}=\frac{G_F}{\sqrt{2}} \frac{\alpha}{2\pi {\tt
sin^2}\theta_W} W_{ds} \cdot \bar{d}\gamma_\mu(1-\gamma_5)s~
\bar{\nu}\gamma^\mu(1-\gamma_5)\nu+H.c.,
\end{eqnarray}
where the short-distance physics is lumped in $W_{ds}$. In the
standard model $W_{ds}$ is dominated by penguin and box diagrams
with intermediate charm and top quark
\begin{eqnarray}
W_{ds}^{SM}=\sum \limits_{l=e,\mu,\tau}~[\lambda_c X_{NL}^l+\lambda_t
X(x_t)~]
\end{eqnarray}
where $ x_t=m_t^2/M_W^2 $, and $\lambda_i=V^*_{is}V_{id}$. The
functions $X_{NL}^l$ and $X(x_t)$ correspond to charm and top
contributions in the loops with the next-to-leading logarithmic
approximation respectively. It has been shown explicitly in Ref.
\cite{Buchalla:1998ba} that the corresponding top contribution can
be written as  $X(x_t)= \eta_t X_0(x_t)$, with the QCD-uncorrected
top quark contribution \cite{{Inami:1980fz}}
\begin{eqnarray}
X_0(x_t)=\frac{x_t}{8}\left[-\frac{2+x_t}{1-x_t}+\frac{3x_t-6}{(1-x_t)^2
}{{\tt ln}(x_t)}\right],
\end{eqnarray}
and the QCD correction factor $\eta_t=0.994$.

Also from table I in Ref.\cite{Buchalla:1998ba}, one has
\begin{equation}
X_{NL}^{e,\mu}=11.00 \times 10^{-4},
\end{equation}
and
\begin{equation}
X_{NL}^\tau =7.47\times 10^{-4}.
\end{equation}
with the central value of the QCD scale $\Lambda =
\Lambda_{\overline{MS}}^{(4)} = 325\pm80 MeV$, and the charm quark
mass $m_c=\bar{m_c}(m_c)=1.30\pm0.05 GeV$.

From the effective Hamiltonian in eq.(3), one can evaluate the hadronic
matrix element $\langle\gamma |J_\mu|K^0\rangle$, where
$J_\mu=\bar{d}\gamma_\mu(1-\gamma_5)s$. Then the corresponding matrix
elements are parameterized as
\begin{eqnarray}
\langle \gamma(q)|d \gamma^\mu \gamma_5 s|K^0(p+q)\rangle &= &-e
\frac{F_A}{M_k}\left[\epsilon^{*\mu}(p \cdot q) -(\epsilon^* \cdot
p)q^\mu\right],
\nonumber \\
\langle \gamma(q)|d \gamma^\mu s|K^0(p+q)\rangle &=& -ie
\frac{F_V}{M_k}\varepsilon^{\mu \alpha \beta
\gamma}\epsilon^*_\alpha p_\beta q_\gamma,
\end{eqnarray}
where $q$ and $p+q$ are photon and $K$-meson four momenta, $F_A$
and $F_V$ are form factors of axial-vector and vector
respectively, and $\epsilon^*$ is the photon polarization vector.
The form factors $F_V$ and $F_A$  have been evaluated in
\cite{Geng:2000ny} using light front quark model.  We find, it is
enough to use momentum independent form factors, $F_A(0)~=~0.0429$
and  $F_V(0)~=~0.0915$ given by the authors of \cite{Geng:2000ny}
to illustrate the numerical estimates in the present paper.

Following the similar way to study $K_{L,S}\to\pi\pi\nu\bar{\nu}$
in Ref. \cite{Chiang:2000bg}, here
 we also consider two representative cases to probe possible new physics
effects of $K_L\to\gamma\nu\bar{\nu}$
 : one is a model-independent way; the other
is a special case--low energy supersymmetry, which however
represents one of the most interesting and consistent extensions
of the standard model.

i) Effective flavor-changing neutral currents (FCNC) interaction.
As formulated by  Nir and Silverman \cite{Nir:1990yq}, it reads
\begin{eqnarray}
{\cal L}^{(z)}=-\frac{g}{4\tt{cos}\theta_W}
     U_{ds}\bar{d}\gamma_\mu(1-\gamma_5)sZ^\mu.
\end{eqnarray}
Combining with the coupling of the $Z$ boson to
neutrino-antineutrino pair, one finds that
\begin{eqnarray}
W^{NP}_{ds}=\frac{\pi^2}{\sqrt{2} G_F M^2_W}U_{ds}=0.93\times 10^2
U_{ds}
\end{eqnarray}
as the new piece of $W_{ds}$ in the effective Hamiltonian
contributing to the basic transition of $s \to d \nu \bar{\nu}$.

The upper bounds on $U_{ds}$ have been determined by other processes
involving $K$-meson and were summarized in Ref. \cite{Nir:1999mg}.
After translating them into the bounds on $W_{ds}$, we obtain
\begin{eqnarray}
|Re(W_{ds})|& \le &0.93 \times 10^{-3}, \nonumber \\
|W_{ds}| & \le & 2.8 \times 10^{-3},  \nonumber \\
|Re(W_{ds})Im(W_{ds})|& \le &1.1 \times 10^{-5}, \nonumber \\
|Im(W_{ds})|& \le &0.93 \times 10^{-3}. \label{UPPER}
\end{eqnarray}
The bound on $|W_{ds}|$ can
be improved using the most recent measurement \cite{Adler:2001xv}
of $K^+ \to \pi^+ \nu \bar{\nu}$, which is
$|W_{ds}|=1.0^{+0.46}_{-0.31} \times 10^{-3}$.

ii) Supersymmetry: Possible interesting supersymmetric effects arise
from penguin diagrams involving charged-Higgs plus top-quark
intermediate states or squark and chargino intermediate states.
These give additional pieces to the effective Hamiltonian of the
form \cite{Colangelo:1998pm}
\begin{eqnarray}\label{SUSY}
W_{ds}^{NP}=\lambda_{t}\frac{m_H^2}{M_W^2 \tt{tan}^2 \beta}H(x_{tH})+
\frac{1}{96} \tilde{\lambda_t},
\end{eqnarray}
where tan$\beta$ is the ratio of the two Higgs vacuum expectation
values and $x_{tH}=m^2_t/M_{H^\pm}^2$. The quantity $H(x)$ is
given by
\begin{eqnarray}
H(x)=\frac{x^2}{8} \left[ -\frac{{\tt ln} x}{(x-1)^2} +\frac {1}
{x-1}\right].
\end{eqnarray}
The definition of $\tilde{\lambda_t}$ is shown in Ref.
\cite{Colangelo:1998pm}. The $\tilde{\lambda_t}$ part in eq.
(\ref{SUSY}), which comes from chargino-squark diagrams, represents the
dominant supersymmetric effect \cite{Colangelo:1998pm} [Only small value
of  tan$\beta$ might enhance the first term in eq. (\ref{SUSY}), which
however is not in favor with the present phenomenologies]. The bound on
$\tilde{\lambda_t}$ can be imposed by similar
considerations  that were used for $W_{ds}$. In Ref.
\cite{Colangelo:1998pm}, the upper limits $|Re \tilde
{\lambda_t}|\le 0.21$ and $|\tilde{\lambda_t}|\le 0.35$ have been
obtained from the observed branching ratios of decays $K_L \to
\mu^+ \mu^-$ and $K^+ \to \pi^+ \nu \bar{\nu}$. Using the most
recent experiment data \cite{Adler:2001xv}, the last limit could
be updated as $|\tilde{\lambda_t}|\le0.14$.

Using the effective Hamiltonian for $K^0 \to \gamma \nu \bar{\nu}$
in eq.(1) together with $K_L \simeq K_2 =
(K^0+\bar{K^0})/\sqrt{2}$, the amplitude of $K_L \to \gamma \nu
\bar{\nu}$ can be divided into the $CP$-conserving and
$CP$-violating parts as follows
\begin{eqnarray}
{\cal M}(K_L \to \gamma \nu \bar{\nu})={\cal M}_{CPC}+{\cal
M}_{CPV}, \nonumber
\end{eqnarray}
where
\begin{eqnarray}
{\cal M}_{CPC}&=&-\frac{G_F}{\sqrt 2}\frac{\alpha}{2\pi \tt{sin}^2
\theta_W} \frac{2}{\sqrt 2}Re(W_{ds})\epsilon^{\mu\nu\alpha\beta}
\frac{F_V}{M_K} \epsilon^*_\mu q_\alpha p'_\beta \bar{u(p_\nu)}
\gamma_\nu (1-\gamma_5)v(p_\nu),
\end{eqnarray}
and
\begin{eqnarray}
 {\cal
M}_{CPV}&=&-\frac{G_F}{\sqrt 2}\frac{\alpha}{2\pi \tt{sin}^2
\theta_W} \frac{2}{\sqrt 2}Im(W_{ds}) \frac{F_A}{M_K}
\epsilon^*_\mu(-p'\cdot q g^{\mu\nu}+ p'^\mu q^\nu) \bar{u(p_\nu)}
\gamma_\nu (1-\gamma_5)v(p_\nu)
\end{eqnarray}
In $K_L$ rest frame the partial decay rate of $K_L \to \gamma \nu
\bar{\nu}$ is given by
\begin{eqnarray}
d^2\Gamma=\frac{1}{(2\pi)^3} \frac{1}{8M_K} | {\cal M} | dE_\gamma
dE\nu.
\end{eqnarray}
By integrating the $E_\gamma$ and $E_\nu$, we obtain
\begin{eqnarray}
\Gamma_{CPC}=\frac{\alpha}{15} \left(\frac{G_F\alpha}{16\pi^2 {\tt
sin^2}\theta_W} \right)^2 |Re(W_{ds})|^2F_V^2M_K^5,
\\
\Gamma_{CPV}=\frac{\alpha}{15} \left(\frac{G_F\alpha}{16\pi^2 {\tt
sin^2}\theta_W} \right)^2 |Im(W_{ds})|^2F_A^2M_K^5.
\end{eqnarray}
Using ${\tt sin}^2\theta_W=0.23$ and $M_K=0.5\ GeV$ the decay
branching ratios are found to be
\begin{eqnarray}
B(K_L \to \gamma \nu \bar{\nu})_{CPC}&=& 6.7\times10^{-8}
|Re(W_{ds})|^2,\\
B(K_L \to \gamma \nu \bar{\nu})_{CPV} &=& 1.4\times10^{-8}
|Im(W_{ds})|^2.
\end{eqnarray}

The decay branching ratios of $K_L\to\gamma\nu\bar{\nu}$, within
the standard model, have been evaluated by Geng, Lih, and Liu
\cite{Geng:2000ny}
\begin{eqnarray}\label{CPCSM}
B(K_L \to \gamma \nu \bar{\nu})^{SM}_{CPC}=1.0\times 10^{-13},
\end{eqnarray}
and
\begin{eqnarray}\label{CPVSM}
 B(K_L\to \gamma \nu \bar{\nu})^{SM}_{CPV}=1.5\times 10^{-15}.
\end{eqnarray}

For the representative scenarios of physics beyond the standard model,
the model-independent upper limits of the $|Re(W_{ds})|$
and $|Im(W_{ds})|$ have been given in eq.(\ref{UPPER}).
 Thus the maximal values of the
branching ratios of the decay could be
\begin{eqnarray}\label{CPCBSM}
 B(K_L \to \gamma \nu \bar{\nu})_{CPC}=1.8\times 10^{-13},
\end{eqnarray}
and
\begin{eqnarray}\label{CPVBSM}
B(K_L \to \gamma \nu \bar{\nu})_{CPV}=0.42\times 10^{-13}.
\end{eqnarray}
Note that a factor 3 has been multiplied in deriving eqs.
(\ref{CPCBSM}) and (\ref{CPVBSM}) since all three flavor neutrinos
should be taken into account. Also from the previous discussions on the
limits of $\tilde{\lambda_t}$, it is easy to see that supersymmetric
extensions of the standard model could allow $|Re(W_{ds})|$ and
$|Im(W_{ds})|$ to reach the upper limit
$0.93\times 10^{-3}$ without conflict with other constraints.

In conclusion, it is shown that the upper limit value of $CP$-conserving
branching ratio in eq.(\ref{CPCBSM}) can be of the same order of
magnitude as the result given in the standard
model, while the upper limit value of $CP$-violating branching ratio in
eq.(\ref{CPVBSM})
is nearly a factor of 30 larger than the value within the standard
model.
Thus it seems that there exists room to look for new
physics beyond the standard model by observing the decay
$K_L\to\gamma\nu\bar{\nu}$ although this is not a very easy task. In
addition to investigations on the
decays of $K\to\pi\nu\bar{\nu}$ and $K\to\pi\pi\nu\bar{\nu}$, it
is therefore expected that both experimental and theoretical
studies of $K_L\to\gamma\nu\bar{\nu}$ could provide complementary
information on the basic weak transition $s\to d \nu\bar{\nu}$.

\begin{center}
{\bf ACKNOWLEDGMENTS}
\end{center}
This work is partially supported by NSF of China 90103002 and 19905008.

\end{document}